\documentclass{iaus}

\usepackage{graphicx}
\usepackage{natbib}
\usepackage{amssymb}
\usepackage{journal_names}

\title[Chemistry in disks]
{Chemical evolution of a protoplanetary disk}

\author[Dmitry A. Semenov]   
{Dmitry A. Semenov$^1$}

\affiliation{$^1$Max Planck Institute for Astronomy, K\"onigstuhl 17, D--69117 Heidelberg, Germany
\\ email: {\tt semenov@mpia.de}}

\pubyear{2011}
\volume{280}  
\pagerange{xxx--yyy}
\jname{The Molecular Universe}
\editors{J. Cernicharo \& R. Bachiller, eds.}
\begin{document}

\maketitle

\begin{abstract}
In this paper we review recent progress in our understanding of the chemical evolution of protoplanetary disks.
Current observational constraints and theoretical modeling on the chemical composition of gas and dust in these systems
are presented. Strong variations of temperature, density, high-energy radiation intensities in these disks, both radially and
vertically, result in a peculiar disk chemical structure, where a variety of processes are active. In hot, dilute and
heavily irradiated atmosphere only the most photostable simple radicals and atoms and atomic ions exist, formed by gas-phase
processes. Beneath the atmosphere a partly UV-shielded, warm molecular layer is located, where high-energy radiation drives rich
ion-molecule and radical-radical chemistry, both in the gas phase and on dust surfaces. In a cold, dense, dark disk midplane
many molecules are frozen out, forming thick icy mantles where surface chemistry is active and where complex polyatomic (organic)
species are synthesized. Dynamical processes affect disk chemical composition by enriching it in
abundances of complex species produced via slow surface processes, which will become detectable with ALMA.
\end{abstract}

\firstsection 
\section{Introduction}
Protoplanetary disks (PPDs) are ubiquitous around young stars \citep[e.g.,][]{LP80,Lissauer_87a}.
Their chemical composition and physical properties regulate
the efficiency and timescale of planet formation. Molecules
and dust serve as heating and cooling agents of the gas, while dust grains also
dominate the disk opacities. Molecular lines observed at infrared and
(sub-)millimeter wavelengths are useful probes of physical conditions (temperature,
density, kinematics, turbulence) in PPDs.

Over the past decade significant progress has been achieved in
our understanding of disk chemical composition, both theoretically and
observationally. Since molecular hydrogen is not observable when it is cold,
we have to rely on other trace species to infer disk physics and chemistry
(Table~\ref{obs_methods}). Multi-molecule, multi-transition interferometric observations, coupled to line
radiative transfer and chemical modeling, allowed to constrain disk sizes,
kinematics, distribution of temperature, surface density, and molecular column
densities (see reviews by \citet{Bergin_ea07} and \citet{DGH07}).

Observations of thermal dust emission are
employed to constrain dust properties, infer disk masses, whereas IR spectroscopy
is used to study mineralogical composition of dust in PPDs. Recently, with space-borne ({\it Spitzer}) and ground-based (Keck,
VLT, Subaru) infrared telescopes,
molecules have been detected in very inner zones of
planet-forming systems, at $r \lesssim 1-10$~AU.

The conditions of planets formation in the early Solar system have been
revealed by a detailed analysis of chemical and mineralogical composition of
meteoritic samples and cometary dust particles
\citep[e.g.,][]{Bradley_05}. The recent {\it Stardust} and {\it Genesis}
space missions have returned first samples of pristine materials, likely of
cometary origin, showing a complex structure of high-temperature crystalline
silicates embedded in low-temperature condensates
\citep{Brownlee_ea04,Flynn_ea06,Brownlee_ea08}. The presence of crystalline silicates
in outer regions of protoplanetary disks has also been revealed
\citep[e.g.,][]{vanBoekel_ea04,Juhasz_ea10a}. An isotopic analysis of refractory condensates in unaltered chondritic
meteorites shows strong evidence that the inner part of the Solar Nebula has
been almost completely mixed during the first several Myr of evolution
\citep[e.g.,][]{Boss2004,Ciesla_09}.

These intriguing findings are partly understood in modern astrochemical models
of protoplanetary disks
\citep{ah1999,Mea02,vZea03,TG07,Agundez_ea08,Woods_Willacy08,Visser_ea09,Walsh_ea10,Semenov_Wiebe11a}.
The major result of the
chemical modeling is that disks have a layered chemical structure
due to heavy freeze-out of gas-phase molecules in the cold midplane and their
photodissociation in the atmosphere. Observed column densities
of CO, HCO$^+$, N$_2$H$^+$, CN, HCN,
HNC, CS, etc. are qualitatively agree with the chemical models.

In this paper we briefly review the major observational findings and modeling predictions for the chemical
composition and evolution of protoplanetary disks.

\section{Observational constrains on composition of gas and ices}

\begin{table}
\caption{Molecular species commonly utilized to study disks}
\label{obs_methods}
\vspace*{1.5ex}
\begin{center}
\begin{tabular}{llcccc}
\hline
{\bf Tracer}              & {\bf Quantity}     & {\bf Midplane}    & {\bf Molecular}    & {\bf Atmosphere}  & {\bf Inner} \\
                          &                     &                  & {\bf layer}  &                         & {\bf zone} \\
\hline
$^{12}$CO, $^{13}$CO      & Temperature        & mm$^{*}$   &  mm     & mm         &  IR \\
H$_2$                     & ---                & 0          &  0      & 0          &  IR \\
NH$_3$                    & ---                & cm         &  cm     & 0          &  0 \\
CS, H$_2$CO               & Density            & 0          &  mm     & 0          &  IR  \\
CCH, HCN, CN              & Photochemistry     & 0          &  mm     & 0          &  IR  \\
HCO$^+$                   & Ionization         & 0          &  mm     & 0          & 0   \\
N$_2$H$^+$                & ---                & mm         &  0      & 0          & 0   \\
C$^+$                     & ---                & 0          &  0      & IR         & IR   \\
complex organics          & Surface            & IR$^{**}$  &  IR-cm  & 0          & IR,mm \\
                          & processes          &            &          &            &     \\
DCO$^+$, DCN,             & Deuterium          & mm         &   mm     &  0         & 0 \\
H$_2$D$^+$                & fractionation      & & & & \\
\hline
\end{tabular}
\end{center}
\begin{flushleft}
$^{*}$ -- ``mm/cm'' and ``IR'' mean radio-interferometric and infrared observations, respectively.\\
$^{**}$ -- Complex molecules frozen onto dust surfaces could be detected through absorption lines
in infrared, while the gas-phase counterparts emit at (sub-) millimeter frequencies.  \\
\end{flushleft}
\end{table}

Apart from CO
and its isotopologues, and occasionally HCO$^+$, DCO$^+$, CN, HCN, DCN, CCH, H$_2$CO, and
CS, the molecular content of protoplanetary disks
remains largely unknown \citep[e.g.,][]{DGG97,Kastner_ea97,Aikawa_ea03,Thi_ea04,Pietu_ea07,Qi_ea08,Henning_ea10}.
Molecular line data are limited in sensitivity and resolution. Thus the spatial distribution of molecular abundances is
still poorly determined
\citep[e.g.,][]{Pietu_ea05,Dutrey_ea07,Panic_ea09}.
Observational facilities such as the Plateau de Bure
interferometer (France) and Submillimeter Array (USA) have permitted measurements of several brightest nearby disks
(DM Tau, LkCa 15, AB Aur, and TW Hya). Soon, with the Atacama Large Millimeter Array entering it's full power in 2013,
we will make substantial progress in our knowledge of molecular disk structure.

Typically, studies of disk physics begin with observations
of bright CO lines. These lines are thermally
excited at densities $\sim 10^3-10^4$~cm$^{-3}$. The $^{12}$CO lines are optically
thick and their intensities measure kinetic temperature in the upper
disk layer \citep[e.g.,][]{DGG97}. The lines of less abundant $^{13}$CO and
C$^{18}$O are typically optically thin or partially optically thick and are
sensitive to both temperature and corresponding column densities
throughout the entire disk. Strong CO lines are most suitable for
accurate determination of disk kinematics as well as orientation and
geometry. Their measured widths indicate that turbulence in disks is
subsonic, with typical velocities of about $0.05-0.2$~km\,s$^{-1}$
\citep{DGH07,Semenov_ea10a,Hughes_ea11a}.

It has been found that disks appear progressively larger from
observations of dust continuum, onwards to C$^{18}$O, $^{13}$CO, and
$^{12}$CO, respectively, with typical values of $300-1000$~AU. This is
a manifestation of selective isotopic photodissociation. Most disks exhibit  a vertical
temperature gradient, ranging from $\sim 10$~K at the midplane to $\sim50-100$~K
in the atmosphere region, as determined by physical models
\citep[e.g.,][]{DDG03,Qi_ea06,Pietu_ea07,Isella_ea10a}, though several disks with large
inner cavities do not show evidence for such a gradient \citep[e.g.,
GM~Aur and LkCa 15;][]{Dutrey_ea08,Hughes_ea09}.

By comparing intensities of 6-5 to 2-1 CO transitions
it was found by \citet{Qi_ea06} that the TW Hya disk has a surface
region that is superheated, so that an additional heating mechanism
is required, possibly stellar X-ray radiation \citep{zetaxa}.
A significant reservoir of very cold CO,
HCO$^+$, CN and HCN gases has been found in the disk of DM~Tau at
temperatures $\lesssim$ 6-17~K, which cannot be explained by conventional chemical
models without invoking a non-thermal desorption or transport mechanism
\citep[e.g.,][]{Semenov_ea06,Aikawa_07,Hersant_ea09}.

The second most readily observed molecular species in
disks is HCO$^+$. The low-lying transitions of this ion are
thermalized at densities of about 10$^5$~cm$^{-3}$. This is one of the most
abundant charged molecule in disks, the other being C$^+$ (not
observable at millimeter wavelengths) and H$_3^+$ (lacking dipole moment).
The ionization degree measured from HCO$^+$ is $\sim 10^{-10}-10^{-9}$ inside warm
molecular disk layer \citep{Qi_ea03,Dutrey_ea07}.

Less strong lines of C$_2$H, CN, and HCN are sensitive to the
intensity and shape of the incident UV spectrum and are excellent
tracers of photochemistry \citep[e.g.,][]{BCDH03}.
Recently detected DCO$^+$ and DCN have abundances that are about $1-10\%$ of the HCO$^+$ and HCN densities
\citep{Qi_ea08}. It remains to be verified whether such a large
degree of deuteration is a heritage of cloud chemistry or produced
{\it in situ}.

A key observational result is that molecular
abundances are depleted by factors 5-100 compared to the
values in the Taurus Molecular Cloud
\citep{DGH07}. Since disks have higher
densities up to $10^7-10^{10}$~cm$^{-3}$ and enshrouded in stronger ionizing
radiation fields, this depletion can be attributed to a combined
effect of photodissociation and freeze-out.

The results from the Infrared Space Observatory and Spitzer telescope
have proven the existence of a significant amount of frozen material
and various types of silicates and polycyclic aromatic hydrocarbons
(PAH) in disks
\citep{vdAea00,vanDishoeck_ARA2004,Bouwman_ea08}. The PAH features at $\sim 3$--12$\mu$m probe the
incident radiation field and density distribution of the upper disk \citep[e.g.,][]{Habart_ea04}.
Recently, with space-borne ({\it Spitzer}) and ground-based (Keck, VLT, Subaru) infrared
telescopes, molecules have been detected in very inner zones of
planet-forming systems, at $r \lesssim 1-10$~AU.
Rotational-vibrational emission lines from CO, CO$_2$, C$_2$H$_2$,
HCN, OH, H$_2$O imply a rich chemistry driven
by endothermic reactions or reactions with activation barriers and photoprocesses
\citep{Lahuis_ea06,Carr_Najita08,Salyk_ea08,Pontoppidan_ea08,Pascucci_ea09,
vdPlas_ea09,Salyk_ea11a}.
Through {\it ISO} and {\it Spitzer} infrared spectroscopy abundant
ices in cold disk regions consisting of water ice and
substantial amounts ($\sim 1-30$\%) of volatile materials like CO, CO$_2$,
NH$_3$, CH$_4$, H$_2$CO, and HCOOH have been detected
\citep[e.g.,][]{Pontoppidan_ea05,Terada_ea07a,Zasowski_ea08}.

In Table~\ref{obs_methods}  the various
molecules used to study protoplanetary disks are summarized.

\section{Global evolutionary picture of disk chemical structure}

\begin{figure}
\includegraphics[angle=0,width=9cm]{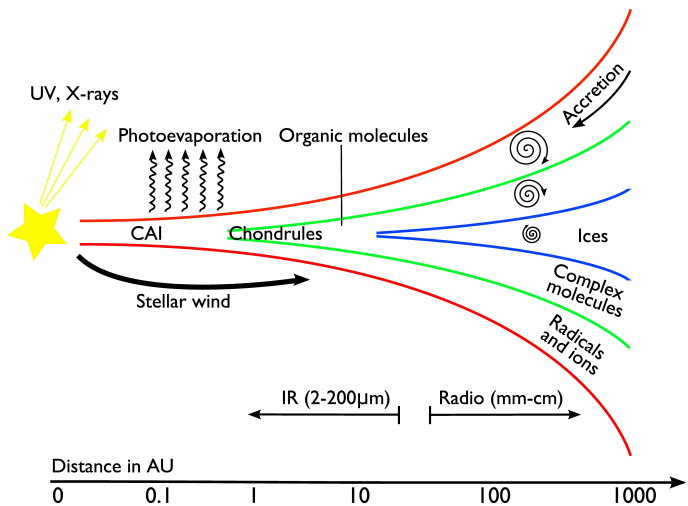}
\caption{Physical and chemical structure of a protoplanetary
disk. Timescales are rough estimates and given for a radius of about
100 AU.
\label{scheme}}
\end{figure}

The current scheme of the disk chemical structure is
presented in Fig~\ref{scheme}.
The disk can be divided into 4 distinct
chemical regions, primarily determined by temperature.  The warm ``inner
zone'' corresponds to radii $< 20$~AU that are accessible with the IR instruments,
while the other 3 regions dubbed ``midplane'', ``molecular layer'', and ``atmosphere'' represent the
outer disk regions ($r  > 20-50$~AU) that are observed with radio-interferometers.

Dense midplane is
opaque to the UV and X-ray radiation and remains cold and essentially neutral ($10-20$~K).
Chemical complexity in this region is
initially reached by fast ion-molecular reactions in the gas,
followed by slow accretion onto grains and  surface reactions (mostly hydrogenation).
Once formed, a molecule is
seldom re-emitted into the gas phase. Typical chemical timescales
for this region are determined by freeze-out and surface reactions,
and are $>10^5-10^6$ years in the outer disk. In the very
inner part, the midplane is hot due to viscous accretion heating ($T\gtrsim 50$~K)
and gas-grain interactions are
not important, so the chemical timescale is due to gas-phase neutral-neutral reactions ($\sim$100 years).
The planet-forming zone is in general in a chemical steady-state, which allows the application of
simple condensation/evaporation thermodynamical chemical models.

Adjacent to the midplane lies a less dense, warmer layer ($T\sim30-70$~K)
that is  partly shielded from stellar and interstellar UV/X-
ray radiation. Young stars emit intense non-thermal UV and thermal
and variable X-ray radiation fields that strongly mediate disk
chemistry at intermediate heights \citep[e.g.,][]{Bergin_ea07}.
UV excess in T Tauri stars is related to their chromospheric activity
\citep[e.g.,][]{Bouvier_ea07}. Soft X-ray radiation has likely the same origin,
while harder X-rays are produced in magnetic reconnection loops or due to jets
\citep[e.g.,][]{Guedel_Naze09}.

These energetic photons dissociate and ionize gas and
photodesorb surface species, thus enriching the gas composition and
initiating a rich chemistry. Abundances of most molecules attain
high concentrations in this zone, and numerous molecular lines are
excited and observable emission produced.
Chemistry does not reach a steady state in this region and a typical
timescale, as determined by surface chemistry and gas-grain interactions,
exceeds $10^5$ years.

Above the intermediate layer a hot, dilute, and heavily
irradiated disk atmosphere exists ($T\gtrsim 100$~K). This is a molecularly-poor region, where only simple light
hydrocarbons, their ions, and other radicals such as CCH and CN
are able to survive. Chemical timescales are short ($\sim$100 years) and defined by
photochemical processes and dissociate recombination.

\section{Gas-phase chemistry}
Models of chemical evolution employ a set
of chemical reactions of relevance to conditions in disks. Modern
astrochemical databases include up to 600 species involved in 4-7000 reactions
\citep{OSU03,Lepetit_ea06,Woodall_ea07,Wakelam_09b}.
Only $10-20\%$ of reaction rates have been studied in the laboratory or calculated
theoretically and thus models are prone to inherent uncertainties
 \citep[see, e.g.,][]{Wakelam_ea10}. All reactions can be divided into  4 distinct groups that are dominant in
different disk regions, see Table~\ref{bond}
\citep[after][]{vanDishoeck88}.

\begin{table}
\caption{Chemical reactions active in disks}
\label{bond}
\vspace*{1.5ex}
\centering
\begin{tabular}{llcccc}
\hline
{\bf Process} & {\bf Formula} & {\bf Midplane} & {\bf Molecular}   & {\bf Atmosphere}   & {\bf Inner} \\
              &         &      & {\bf layer} &    & {\bf zone} \\
\hline
{\bf Bond formation} & & & & & \\
Radiative association  & A + B $\rightarrow$ AB + h$\nu$          & X         &  X    & X   &  X  \\
Surface formation      & A + B$\|$gr    $\rightarrow$ AB + gr    &  X         & X    & 0   &  0 \\
Three-body             & A + B + M $\rightarrow$ AB + M          &  0         & 0    & 0   & X \\
\hline
{\bf Bond destruction} & & & & & \\
Photodissociation      & AB + h$\nu$ $\rightarrow$ A + B          &  0         & X    & X   & X \\
Dissociation by CRP & AB + CRP $\rightarrow$ A + B                & X           & X    & 0    & 0 \\
Dissociation by X-rays & ---                                      & 0           & X    & X    & X \\
Dissociative           & AB$^+$ + e$^-$ $\rightarrow$ A + B       & X          & X    & X   & X  \\
recombination          & & & & & \\
\hline
{\bf Bond restructuring} & & & & & \\
Neutral-neutral &      A + BC $\rightarrow$ AB + C                & X         & X     & 0   & X \\
Ion-molecule          & A$^+$ + BC $\rightarrow$ AB$^+$ + C      &  X         & X     & X   & X \\
Charge transfer        & A$^+$ + BC $\rightarrow$ A + BC$^+$     &  X         & X     & X   & X \\
\hline
{\bf Unchanged bond} & & & & & \\
Photoionization         & AB + h$\nu$ $\rightarrow$ AB$^+$ + e$^-$ & 0       & X     & X   & X \\
Ionization by CRP & AB + CRP $\rightarrow$ AB$^+$ + e$^-$           & X       & X     & 0   & 0 \\
Ionization by X-rays    & ---                                       & 0       & X     & X   & X \\
\hline
\end{tabular}
\end{table}


Apart from the very dense inner zone, all reactions in disks are
two-body processes. Three-body reactions become competitive only
at $\lesssim 10$~AU, where $n \gtrsim 10^{10}$~cm$^{-3}$ \citep{Aea99}.
The main processes leading to formation of molecular bonds are slow radiative
association and surface reactions.
Upon collision, a collisional complex in an excited state may
form, which is stabilized with a low probability by emission of a
photon \citep[e.g.,][]{Bates51,Williams72,HerbstKlemperer73}.
For example, formation of light hydrocarbons
starts with radiative association of C$^+$ and H$_2$,
leading to excited CH$_2^+$ \citep[][]{Herbst85}.

Ionized by FUV, X-ray photons or CRPs, ions and molecule drive rapid ion-neutral chemistry, which constitutes the
largest fraction of astrochemical models. These reactions are
exothermic, with high rate coefficients $\sim 10^{-9}$~cm$^{-3}$\,s$^{-1}$,
which often increase toward low temperatures ($\beta<0$) \citep[e.g.,][]{DalgarnoBlake76}.
Ion-neutral reactions result in bond restructuring of
the reactants. Some of the most important reactions of this category are protonation reactions.

Molecular ions are destroyed by
dissociative recombination with electrons and negatively charged grains. These
processes are especially fast at low temperatures, with typical rates
of about $10^{-7}$~cm$^{-3}$\,s$^{-1}$ at $10$~K \citep{Woodall_ea07}. For nearly all
observed species, dissociative recombination is an important
formation pathway (e.g., water and hydrocarbons). Often, at later
evolutionary times, $\gtrsim 10^5$~years, dissociative recombination is balanced by
protonation reactions, e.g. CO + H$_3^+$ $\rightarrow$ HCO$^+$ + H$_2$ followed by HCO$^+$ + e$^-$ $\rightarrow$ CO + H.
Products and branching ratios for polyatomic ions are
not easily obtainable \citep[e.g.,][]{BatesHerbst_inratecoeff88,SmithSpanel94}.

A number of neutral-neutral reactions involving open-shell radicals
can also be active in both cold outer and warm inner disk regions \citep{vanDishoeckxx}.
The typical rate coefficient for these reactions is
$\sim10^{-11}$--$10^{-10}$~cm$^{-3}$\,s$^{-1}$, i.e., only
about an order of magnitude lower than for the ion-molecule
processes \citep[e.g.,][]{Clary85,OSU03}. One of the
most interesting reactions of this type is formation of HCO$^+$ upon
collision between O and CH \citep[$\alpha_0=2.0\cdot 10^{-11}$~cm$^{-3}$\,s$^{-1}$;][]{Woodall_ea07}.


\section{Photochemistry}
The photochemical reactions are overviewed in this book in the paper by van Dishoeck \& Visser and only
briefly discussed here.

Young T Tau stars ($T_{\rm eff} \simeq 4\,000$~K) emit intense non-thermal UV radiation
that has a spectra different from the interstellar UV field \citep{BCDH03}, with a prominent Lyman$_\alpha$
line,
while hot ($T_{\rm eff} \gtrsim 10\,000$~K) Herbig Ae/Be
stars produce a lot of thermal UV emission.
The overall intensity of the stellar UV radiation at 100~AU from the star can be as
high as 500 and $10^5$ for a T Tau and a Herbig Ae star, respectively, in units of the interstellar UV field
\citep{Habing68}.

Many molecules, such as CO, H$_2$, and CN, are dissociated by
radiation at short wavelengths ($\lambda \lesssim 1\,100$~{\AA}), while
photodissociation of other species, such as HCN, occurs at longer
wavelengths \citep[][]{vanDishoeck88,vDea_06}.
Some molecules (e.g., H$_2$ and CO) dissociate
via absorption of UV photons at discrete
lines to the excited (Rydberg) states, whereas other molecules are
dissociated either by the continuum (e.g., CH$_4$) or by the
continuum and lines (e.g., C$_2$) \citep[][]{vD88,vDea_06}.
The high ratio of CN to HCN abundances, observed in disks \citep{DGG97}, can be explained by more
intense photodissociation of HCN if part of the UV flux comes as Ly$_\alpha$ photons
\citep[$1216\AA{}$;][]{BCDH03}.

Since the dissociation of the abundant H$_2$ and CO molecules
result from photoabsorption at discrete wavelengths, isotopically
selective photodissociation based on self-shielding is possible \citep[e.g.,][]{Visser_ea09b}.
Two conditions for that are required:
1) dissociation via line absorption for each isotopically
substituted molecule, and 2) differential photolysis that depends
upon the isotopic abundances. Self-shielding occurs
when the spectral lines leading to dissociation of the major isotopic species
optically saturate, while the other residual lines relevant for dissociation of the minor isotopes
remain transparent. Under conditions of protoplanetary disks,
isotopic-selective chemistry may occur for abundant CO molecules
\citep[e.g.,][]{DS70,Thiemens1983,vanDishoeck88,Lea96,Visser_ea09b}.
%

\section{Gas-grain interactions}
In cold disk regions grains
serve as a passive sink for heavy molecules and electrons, while
providing free electrons to dissociate positive ions. In the
darkest midplane zone dust grains become the dominant charged species
\citep{Red2}.
At low temperatures of $\lesssim 20-50$~K, many molecules adhere to a
grain with a nearly 100\% sticking probability since their kinetic
energy is much smaller than the binding energy \citep[see, e.g.,][]{dHendecourtea85,BuchZhang91}.
This sticking
probability also depends on surface properties, such as porosity and
distribution of surface chemi- and physisorption sites.
Chemisorption requires
formation of a chemical bond between a surface species and a grain,
and such a species does not evaporate easily \citep[e.g.,][]{Cazaux_ea05}.

The most effective desorption processes for disk chemistry are
thermal evaporation, cosmic ray induced desorption, and
photodesorption. Thermal evaporation occurs if a molecule has
energy that exceeds it's binding energy. Typical binding energies of
physisorbed species are about 1\,000~K for light molecules like CO and N$_2$
\citep{Bisschop_ea06}, and much larger for
heavier cyanopolyynes and carbon chains. Chemisorbed species do not
desorb until very high temperatures of $100-1000$~K are reached.

In disk midplane CRPs provide energy that allows dust mantles (partly) thermally evaporate.
A relativistic iron nucleus may eventually collide
with a grain and impulsively heat it, releasing a portion of
the volatile component \citep[e.g.,][]{WatsonSalpeter72,Leger_ea85,HW90}.
In less opaque disk regions, penetrating UV photons lead to
photoevaporation of surface species. The probability of evaporation
per one UV photon has been measured in the laboratory for some
simple molecules, such as CO, H$_2$O, CH$_4$, and NH$_3$, and is
about $10^{-6}$--$10^{-2}$ \citep[e.g.,][]{Oeberg_ea07,Oeberg_ea09a,Oeberg_ea09b}.
The dilute UV radiation field produced
by cosmic rays in the disk midplane \citep{PrasadTarafdar83}
can also be important for high values of the photodesorption rates.
Another mechanism is the X-ray induced desorption which is efficient only for tiny grains,
$\lesssim0.05\mu$m \citep{Najita_ea01}.

\section{Surface formation of complex species}
The surfaces of dust grains serve as a catalyst for reactions
that do not proceed efficiently in the gas phase.
The most notable example is formation of molecular
hydrogen, which occurs almost entirely on dust surfaces
\citep[e.g.,][]{HollenbachSalpeter71,WatsonSalpeter72}.
\citet{TielensHagen82} have studied the chemical evolution of surface species on large grains and considered the effect
of various migration rates for light atoms and heavy molecules. A typical $0.1\mu$m amorphous silicate
grain accommodates $\sim 10^6$ surface sites available for accretion. An atom or light
radical, if it is not chemisorbed, may migrate over the surface from
site to site by thermal hopping, when its energy exceeds the barriers
for particle motions and react with other species.

There is an increasing
body of evidence from laboratory measurements that complex
molecules like methanol cannot be produced in the gas phase via
radiative association forming a large protonated precursor followed by its dissociative recombination
\citep{Geppert_ea05}. Thus,
surface reactions due to thermal hopping remain the only viable
formation pathway for production of complex (organic) molecules
in protoplanetary disks, found also in meteoritic samples in Solar
system.

\section{Importance of dynamical processes for disk chemistry}
While most of the chemical studies are still based on laminar disk models, evidences for mixing
call for a more sophisticated treatment. Models of the early Solar nebula with radial
transport by advective flows have been developed
\citep[e.g.,][]{Morfill_Voelk84,G01,G02,Wehrstedt_Gail02,Boss2004,Keller_Gail04}.
\citet{IHMM04} for the first time modeled the influence of turbulent diffusion in the vertical direction and
advection flows in the radial direction on the chemical composition of the
inner disk region. They found that dynamical processes significantly
affect the chemical evolution of sulfur-bearing species. \citet{Willacy_ea06}
have shown that 1D vertical mixing modifies chemical composition of the outer
disk region and that the mixing results better agree to observations.
\citet[][Paper~I]{Semenov_ea06} and \citet{Aikawa_07} have found that turbulent transport
allows explaining the presence of a large amount of cold ($\lesssim15$~K) CO gas in the
disk of DM Tau.
\citet{TG07} have used a 2D disk chemo-hydrodynamical model and showed that in
the disk midplane matter moves outward, carrying out the angular momentum, while
the accretion flows toward the star are located at elevated
altitudes. Consequently, gas-phase species produced by warm chemistry in the inner
nebula can be steadily transported into the cold outer region and freeze out.
A radial advection model has also been utilized by \citet{Nomura_ea09}, who have
demonstrated that inward radial transport enhances abundances of organic
molecules (produced mainly on dust surfaces in cold outer regions).
\citet{Hersant_ea09} have studied various mechanisms to retain gas-phase
CO in very cold disk regions. They concluded that efficient photodesorption
in moderately obscured disk regions ($A_{\rm V}<5^{\rm m}$) greatly enhances gas-phase CO
concentrations, while the role of vertical mixing is less important.
Finally, \citet{Heinzeller_ea11} have investigated the disk chemical evolution with
radial advection, vertical mixing, and vertical wind transport processes.  They have found that the disk wind has
a negligible effect on disk chemistry, whereas the radial accretion alters the molecular abundances
in the cold midplane, and the vertical turbulent mixing affects the chemistry in the warm molecular layer.

We study the influence of turbulence mixing on the chemical evolution of protoplanetary disks
in \citet{Semenov_Wiebe11a}. Mixing is important in disks since a chemical steady-state is not
reached due to long timescales associated with surface chemical processes and slow evaporation
of heavy molecules.
Our analysis was based on the $\alpha$-model of a $\sim5$~Myr DM~Tau
disk coupled to the large-scale gas-grain chemical code ``ALCHEMIC'' \citep{Semenov_ea10}. To account for
production of complex molecules, our chemical network was supplied with
a set of surface reactions and photoprocessing of ices.

The adopted flaring disk structure is based on a 1+1D steady-state $\alpha$-model similar
to \citet{DAea99} model. The non-thermal FUV radiation field from DM Tau is
represented by the scaled ISRF of \citet{G},
with the un-attenuated intensity at 100~AU of $\chi_*(100)=410$ \citep[e.g.,][]{Bergin_ea04}.
For the X-ray luminosity of the star we adopt a value of $10^{30}$~erg\,s$^{-1}$, which is constrained by
recent measurements with Chandra and XMM in the range of 0.3-10~keV.

The turbulence in disks is likely driven by the magnetorotational instability (MRI), which is operative even in a
weakly ionized medium \citep[e.g.,][]{MRI}. This turbulence causes anomalous viscosity that
enables efficient redistribution of the angular momentum. We have
followed the parametrization of \citet{ShakuraSunyaev73}, where turbulent viscosity $\nu$ is related to local disk
properties such as the vertical spatial scale $H(r)$, the sound speed $c_{\rm s}(r,z)$,
and the dimensionless parameter $\alpha$:
\begin{equation}
 \nu(r,z) = \alpha\,c_{\rm s}(r,z)\,H(r).
\end{equation}
From observational constraints $\alpha$ is $\sim 0.001-0.1$ \citep[][]{Andrews_Williams07,Guilloteau_ea11a},
so we adopt the constant value of $0.01$. Consequently, the diffusion coefficient is calculated as
\begin{equation}
D_{\rm turb}(r,z) = \nu(r,z)/Sc,
\end{equation}
where $Sc$ is the Schmidt number describing efficiency of turbulent diffusivity \citep[see
e.g.][]{ShakuraSunyaev73,SchraeplerHenning04}.
We assume that gas-phase species and dust grains are well mixed, and transported with the same diffusion coefficient.

\begin{table}
\caption{Detectable tracers of turbulent mixing.}
\vspace*{1.5ex}
\centering
  \label{tab:tracers}
  \begin{tabular}{ll}\hline
{\bf Steadfast} & {\bf Hypersensitive}\\ \hline
CO          &  Heavy hydrocarbons (e.g.,C$_6$H$_6$) \\
H$_2$O ice   &  C$_2$S \\
       &  C$_3$S \\
       &  CO$_2$ \\
       &  O$_2$ \\
      &  SO \\
      &  SO$_2$\\
      &  OCN \\
      &  Complex organics (e.g., HCOOH)\\
\hline
 \end{tabular}
\end{table}

We show that the higher the ratio
of the characteristic chemical timescale to the turbulent transport timescale
for a given molecule, the higher the probability that its column density
will be affected by dynamical processes. Thus, turbulent transport
influences abundances of many gas-phase species and
especially ices. Vertical mixing is more important as it affects the evolution of gas-phase and surface species of any kind,
whereas the effect of radial mixing is pronounced mostly for the evolution of ices.
The radial temperature gradient is weaker, and thus is only
relevant for the evolution of polyatomic ices formed via surface reactions of heavy radicals, whereas steep
vertical gradients of temperature and high-energy radiation intensity
cause much sharper transition from the ice-dominated chemistry in the disk midplane to the rich gas-phase chemistry
in the molecular layer.

The simple molecules that are
unresponsive to dynamical transport include C$_2$H, C$^+$, CH$_4$, CN, CO, HCN, HNC, H$_2$CO, OH,
as well as water and ammonia ices. The sensitive species to dynamics are carbon chains and other heavy species,
in particular sulfur-bearing and complex organic molecules frozen
onto the dust grains. Mixing steadily transports ice-coated grains in
warmer regions, allowing more efficient surface processing due to enhanced hopping rates of heavy
radicals. In warm intermediate layer these organically-rich ices  evaporate, and in the inner disk they can also be
photodissociated by CRP/X-ray-induced UV photons. The importance of
mixing is higher in an inner, planet-forming disk zone, where thermal, density, and high-energy radiation gradients
are stronger than in the outer region. Still, mixing does not completely erase the
layered chemical structure of protoplanetary disks.

Several promising detectable tracers of
dynamical processes in protoplanetary disks are the column density ratios of the CO$_2$,
O$_2$, SO, SO$_2$, C$_2$S, C$_3$S to that of CO and the water ice (see Table~\ref{tab:tracers}).
The detection of complex species
(e.g., dimethyl ether, formic acid, methyl
formate, etc.) in protoplanetary disks with ALMA and JWST will be a strong indication
that chemical evolution of these objects is influenced by transport processes.


%
%

\bibliographystyle{aa}
\bibliography{references}

\begin{discussion}

\discuss{Paul M.~Woods}{There were some dynamical models recently that show that the accretion flow proceeds inwards along the
disc surface and outwards along the disc midplane. Are those models credible, and how would that affect the disc chemistry?}

\discuss{D.~Semenov}{Turbulence is a 3D phenomenon and its magnitude varies greatly across a disk.
Global 3D MHD models show that viscous transport occurs in all directions in any given disk location, albeit with different
efficiencies, and that there is no a specific direction for it. According to observations,
advective transport seems to be an effective mechanism that regulates overall lifetime
of disks, $\sim 1-10$~Myr, in which most of disk matter accretes onto the central star and photoevaporates.
Consequently, disk structure changes with time, as well as the FUV and X-ray penetration, and grain
properties. This strongly affects chemical evolution of protoplanetary disks, and favors the use of evolutionary disk
models (see R.~Vissier's presentation and proceeding paper.}

\discuss{James R.~Lyons}{What turbulent viscosity parameter $\alpha$-values were used for slow vs. fast mixing?}

\discuss{D.~Semenov}{We considered Fickian diffusion, in which the diffusion coefficient $D_{\rm turb}(r,z)$ is attributed to the
$\alpha$-parameterized viscosity, $\alpha=0.01$: $D_{\rm turb}(r,z) = \nu(r,z)/Sc=\alpha\,c_{\rm s}(r,z)\,H(r)/Sc$,
where the characteristic spatial scale is $H(r)$ and the sound speed is $c_{\rm s}(r,z)$. The $Sc$  the
Schmidt number describing efficiency of turbulent diffusivity. The ``fast'' mixing model has $Sc=1$ and the ``slow'' model has
``Sc=100''.}

\discuss{James R.~Lyons}{Have you included isotopes in vertical mixing calculations?}

\discuss{D.~Semenov}{No.}

\discuss{James R.~Lyons}{What did you mean by ``isotopic homogeneity of the Solar nebula''? It is not homogeneous in oxygen
isotopes.}

\discuss{D.~Semenov}{I've meant that the inner, 1-20~AU region of the Solar
nebula has been isotopically homogenized at bulk level, though oxygen anomalies at percentage level still persist.}
\end{discussion}

\end{document}